\documentclass[aps,showpacs,amsmath,amsfonts,amssymb,twocolumn]{revtex4-2}
\usepackage{graphicx}
\usepackage{epsf}
\usepackage{epstopdf}
\usepackage{setspace}
\usepackage{array}
\usepackage{natbib,hyperref}
\begin{document}

\begin{spacing}{1}

\pagenumbering{arabic}

\title{Tunneling and Revival of Anderson Localization in Bose-Einstein Condensate}

\author{Sriganapathy Raghav}
\author{Barun Halder}
\author{Pradosh Basu}
\author{Utpal Roy}
\email{uroy@iitp.ac.in}
\affiliation{Indian Institute of Technology Patna,
Bihta, Patna-801103, India}

\begin{abstract}
We provide an analytical model to fabricate an exponential localization of a Bose-Einstein condensate under bichromatic optical lattice. Such localization is famously known as Anderson localization. The degree of localization is investigated by the Participation Ratio to recognize the laser parameter domain for Anderson localization. The exponential nature of the localization is proved, where we also identify the Localization Length. The tunneling of Anderson-localized condensate with time is observed, and the revival phenomenon of Anderson localization is reported. Slowing down of Anderson localization is noticed for higher laser intensity. We also study the dynamical and structural stability of the condensate during Anderson localization, which suggests the preferred values of laser power and time instance to encounter minimal mean difference in the presence of noise.
\end{abstract}

\maketitle

\section{Introduction}

In 1958, Anderson demonstrated exponential localization when the on-site energies of the lattice get randomly shuffled \cite{anderson1958absence}. Since then, there has been an extensive search for situations in condensed matter systems to manifest exponentially localized wavefunctions in spatial coordinate \cite{lee1985disordered,MOTT19681,abrahams1979scaling,belitz1994anderson}. However, the condensed matter system did not transpire as a favorable candidate, and scientists also started looking for possibilities in ultracold atoms. Billy \emph{et al.} were the first to report exponential localization in ultracold matter waves \cite{billy2008direct}, which is followed by various other groups \cite{roati2008anderson,schwartz2007transport,modugno2010anderson,fallani2008bose,jendrzejewski2012three}. In recent work, Schneider \emph{et al.} observed Anderson localization for a 2D quasi-crystalline optical lattice \cite{sbroscia2020observing}. The study of ultracold atoms in quasi-crystalline optical lattices like bichromatic optical lattice (BOL), has become the center of attraction for quantum simulation research \cite{szpak2011quantum,long2015tunable,thomas2017quantum,kundu@2022}. When two optical lattices of different wavelengths and depths are overlapped, one may obtain a geometrically frustrated potential BOL \cite{schulte2005routes}. A variety of interesting physics has emerged using ultracold atoms trapped in a BOL potential \cite{lucioni2011observation,struck2011quantum,nascimbene2012experimental,salger2007atomic,poletti2008dynamics}. There are numerous theoretical studies about ultracold atoms in BOL by utilizing Bose-Hubbard model, and numerical solutions of Gross-Pitavskii equation (GPE) \cite{modugno2009exponential,shiozaki2011transition,seman2011route,dutta2015non,bhat2021modulational}. Adhikari \emph{et al.} carried out numerical studies on the localization of Bose-Einstein condensate (BEC) in BOL of 1,2 and 3 dimensions with different non-linearities \cite{adhikari2009localization}. The exact analytical model for BEC inside a BOL was reported by Nath \emph{et al.} \cite{nath2014bose,nath2020exact}, where the solutions for both attractive and repulsive nonlinearities were studied with precise trap engineering.

Anderson localization in BEC is thought of as a highly localized condensate cloud of an exponentially varying profile that is formed inside a disordered potential like BOL with periodic lattice frustrations. However, Anderson localization happens only in the frustrated lattice sites instead of other potential minima for a highly intense BOL trap. Identifying a precise parameter regime of Anderson localization is definitely a matter of concern in experiments. A natural question is if this sharply localized cloud in the frustrated site can be transported to the other frustrated sites by overcoming the high potential barrier of the intense BOL. Secondly, during the transportation, if any, what may be the temporal dynamics of the Anderson-localized condensate and whether such localization disappears forever after tunneling.

In this work, we address all these issues with an exact analytical, time-dependent model for BEC under BOL potential. We have proven the exponential localization, for the first time, through an analytical approach and predicted the localization length for BEC. The trap engineering of such localization is also reported for experimental feasibility. The parameter regime of Anderson localization is demarcated by the nature of the participation ratio (PR). Tunneling of such Anderson-localized cloud is studied for the condensate in the presence of an initial drift \cite{schmid2006long,sadgrove2007rectified,groiseau2018quantum}. The nature of the tunneling is illustrated by time snapshots of condensate density and the rise/decay of localization inside the frustrated site. The periodic occurrence of Anderson localization with time is observed and reported as the revivals of such localization during its tunneling.
Moreover, we demonstrate the variation of the revival time with a tunable BOL-parameter and deceleration of the exponentially localized cloud. The numerical stability of the exponentially-localized condensate is studied at different times, establishing the present model as a sufficiently stable configuration.

\section{Analytical Model for Localization}

The dynamics of BEC are analytically modeled using GPE, a modified form of Schr{\"{o}}dinger's equation with a nonlinearity term \cite{pethick2008bose,atre2006class}. The nonlinearity can be controlled by modulating the atomic scattering length using the Feshbach resonance technique. The exact analytical model for BEC has been obtained for various external traps like harmonic, double well, periodic, and quasi-periodic \cite{serkin2000novel,liang2005dynamics,yan2012matter,bronski2001bose}.
Our present work focuses on obtaining the time-varying solution to 1D-GPE for a stationary BOL and studying the transport of Anderson localization. We obtain solutions that cover a wide range of situations, such as periodic/localized, based on modulus parameters of the elliptic function. The bright/dark solitary excitations are obtained based on inter-atomic interactions, and we study how they propagate across the lattice. For building the exact analytical model, we start with a dimensionless 1D-GPE of weakly interacting ultracold atoms with cubic nonlinearity:
\begin{widetext}
\begin{eqnarray}
 i\frac{\partial \psi(z,t)}{\partial t}+\frac{1}{2}\frac{\partial^2 \psi(z,t)}{\partial z^2}-g(z,t)|\psi(z,t)|^2\psi(z,t)-V(z)\psi(z,t)-i\tau(z,t) \psi(z,t)=0.\label{GP}
\end{eqnarray}
\end{widetext}
Here, $\psi(z,t)$ is the 1D condensate wavefunction, $g(z,t)$ denotes the space- and time-modulated nonlinearity, $\tau(z,t)$ represents the space- and time-modulated loss or gain of the condensate atoms. For the purpose of illustration, we have exploited experimentally feasible parameters of $Li^7$-BEC trapped in a 1D-waveguide of transverse frequency $\omega_\perp$=$2\pi\times710$ $Hz$, optical lattice wavelength $\lambda=10.62\mu m$, the $s$-wave scattering length $a_s=-0.21nm$ and the harmonic oscillator length in the transverse direction is $a_\perp=\sqrt{\frac{\hbar}{M\omega_\perp}}=1.4245$ $\mu m$ \cite{khaykovich2002formation}. The dynamics of the condensate can efficiently be controlled by the axial external trap ($V(z)$), which can be engineered by varying the applied magnetic field and the angle between the overlapping laser beams \cite{inouye1998observation}. For obtaining an Anderson localization, one requires a BOL, which possesses frustrated potential sites. A BOL potential is created by the application of two laser beams of different wavelengths and different powers. The potential under consideration is given by
\begin{eqnarray}
  V(z)=V_1\cos(2lz)+V_2\cos(lz).
\end{eqnarray}

The potential depths $V_1$ and $V_2$ are related to the recoil energy for the constituent lasers of wavelength $\lambda$.
The exact spatially localized solution of the variable coefficient nonlinear Schr\"{o}dinger equation (Eq.~(\ref{GP})) is nontrivial in the presence of a trap, and thus one performs similarity transformation, which connects it to a known solvable nonlinear equation with constant coefficients, leaving behind a set of consistency conditions. The corresponding ansatz is written as
\begin{eqnarray}
 \psi(z,t)=A(z,t)F[Z(z-vt)]e^{i\theta(z,t)},\label{ansatz}
\end{eqnarray}
such that, the travelling coordinate is taken in a simple form: $Z(z-vt)=B(z)+D(t)$. Here, $v$ is a uniform speed, which can be associated with experimental situations where the BEC is moved horizontally, either by coherently moving the lattice \cite{schmid2006long}, or imparting momentum by pulsed standing wave laser field \cite{sadgrove2007rectified,groiseau2018quantum}. The condensate density $F[Z(z-vt)]$, and the phase $\theta(z, t)$ are expected to  depend on the trap parameters of the external BOL potential and initial velocity of the cloud. The real part of the equation, upon substituting the above ansatz with travelling coordinate, becomes
\begin{widetext}
\begin{equation}
\begin{aligned}
     -A(z,t)F[Z(z-vt)]&\theta(z,t)_t +\frac{1}{2}A_{zz}(z,t)F[Z(z-vt)]+A(z,t)_z\frac{\partial F[Z(z-vt)]}{\partial Z(z-vt)}\frac{\partial Z(z-vt)}{\partial z}\\
    & +\frac{1}{2}A(z,t)\frac{\partial^2F[Z(z-vt)]}{\partial Z(z-vt)^2}\Bigg(\frac{\partial Z(z-vt)}{\partial z}\Bigg)^2+\frac{1}{2}A(z,t)\frac{\partial F[Z(z-vt)]}{\partial Z(z-vt)}\frac{\partial^2Z(z-vt)}{\partial z^2}\\
    & -\frac{1}{2}A(z,t)F[Z(z-vt)]{\theta(z,t)_z}^2-g(z,t)A(z,t)^3F[Z(z-vt)]^3-V(z)A(z,t)F[Z(z-vt)]=0.
    \end{aligned}
\end{equation}
\end{widetext}
The subscripted variables imply partial differentiation. To explore the connection between trap parameters and the nonlinearity, we have mapped the dynamics with the following known nonlinear differential equation, which also helps us find the proper consistency conditions.
\begin{equation}
    \frac{\partial^2 F[Z(z-vt)]}{\partial Z(z-vt)^2}
    -G F[Z(z-vt)]^3=0,
\end{equation}
where the constant, $G=2g(z,t)A^2(z,t)/Z_z^2 (z,t)$, is the nonlinearity constant. $G$ is $-1$ for attractive and $1$ for repulsive inter-atomic interactions. This nonlinear equation is taken in the form of the elliptic equation, which is known to manifest solitary wave excitations \cite{raju2005exact}. The elliptic equation has $12$ solutions as Jacobi elliptic functions, among which $cn[Z(z-vt),m]$ and $sn[Z(z-vt),m]$ are widely used. The modulus parameter $m$ controls the nature of excitations, such that $m=0$ manifests periodic solution, $0\leq m \leq 1$ produces cnoidal waves, and $m=1$ manifests localized solution. Latter is the case under consideration. After connecting with this elliptic equation, the real part is left with the following consistency conditions, relating the amplitude, phase, and spatial part of the travelling coordinate:
\begin{equation}
    \frac{A_{zz}(z,t)}{2A(z,t)}-\theta_t(z,t)-\frac{{\theta_z(z,t)}^2}{2}-V(z)=0,\label{consistency for potential}
\end{equation}
\begin{equation}
    \Bigg[A^2(z,t)B_z(z)\Bigg]_z=0.
\end{equation}
We carry on a similar calculation for the following imaginary part of Eq.~(\ref{GP}) after the substitution of the ansatz with travelling coordinate:
\begin{widetext}
\begin{equation}
\begin{aligned}
    A_t(z,t)F[Z(z-vt)]+&A(z,t)\frac{\partial F[Z(z-vt)]}{\partial Z(z-vt)}\frac{\partial Z(z-vt)}{\partial t}+A(z,t)_zF[Z(z-vt)]\theta_z(z,t)\\
    &+A(z,t)\frac{\partial F[Z(z-vt)]}{\partial Z(z-vt)}\frac{\partial Z(z-vt)}{\partial z}\theta_z(z,t)
    +\frac{1}{2}A(z,t)F[Z(z-vt)]\theta_{zz}(z,t)\\
    &-\tau(z,t) A(z,t)F[Z(z-vt)]=0,
    \end{aligned}
\end{equation}
\end{widetext}
which brings out another set of consistency equations as follows
\begin{eqnarray}
D_t(t)+B_z(z)\theta_z(z,t)=0,\label{consistency for travelling coordinate}\nonumber\\
\\
2A(z,t)A_t(z,t)+[A^2(z,t)\theta_z(z,t)]_z-2\tau(z,t)A^2(z,t)=0.\nonumber
\\
\end{eqnarray}

\begin{figure}
    \centering
    \includegraphics[width=9cm]{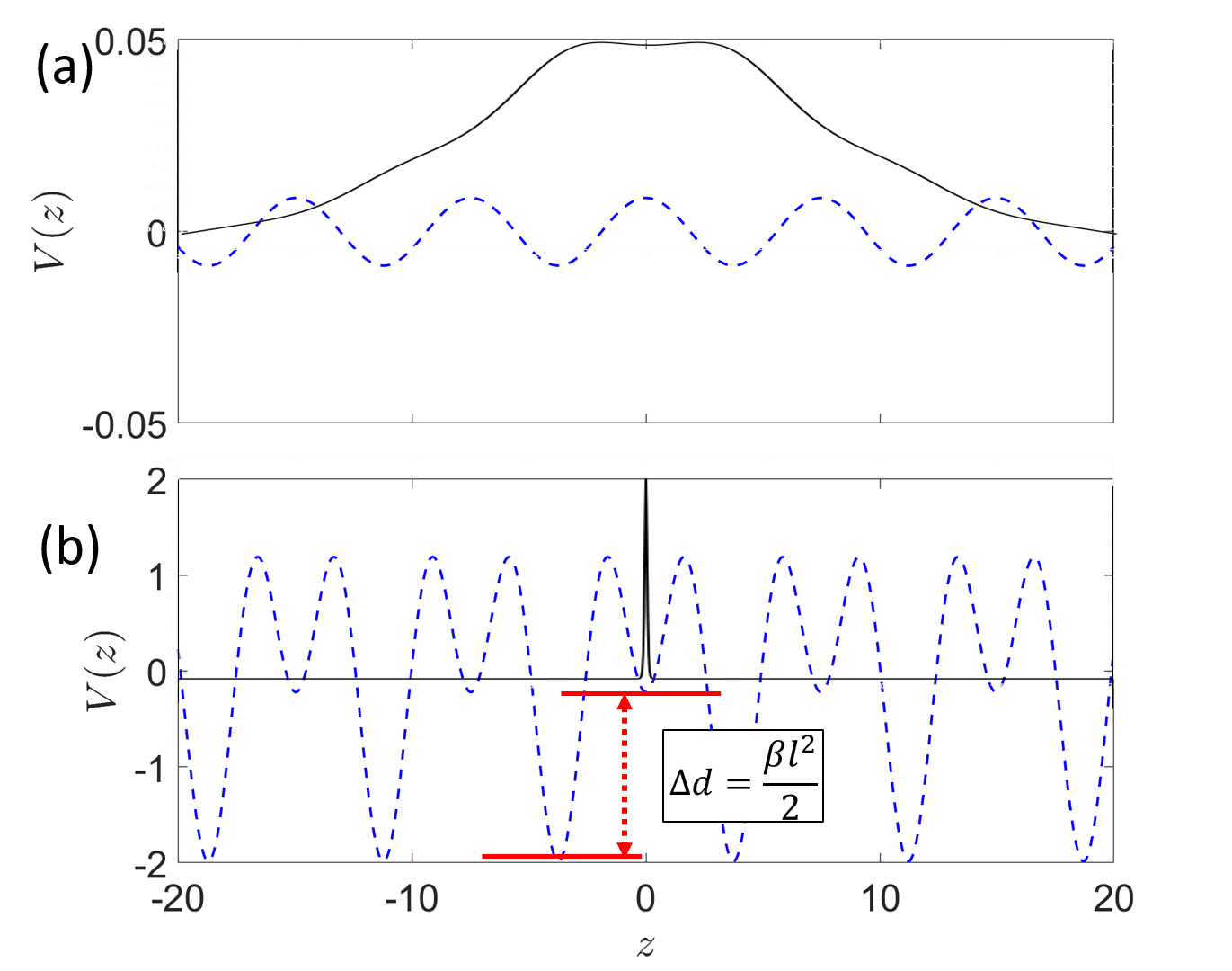}
    \caption{The potential profile (dotted lines) of BOL for varied potential parameters: a) $\beta=0.05$ and b) $\beta=5$. Solid lines represent condensate density.$z$ is scaled by harmonic oscillator length $a_{\perp}=1.4245$ $\mu m$ and $V(z)$ is scaled by $\hbar \omega_{\perp}$.}
    \label{fig:potential}
\end{figure}

The consistency equations are carefully solved to obtain the following solutions, subject to the substitution of the form of the BOL trap:
\begin{eqnarray}
D(t)=-vt+a_1, \;\;\; \theta_z(z,t)=\frac{v}{B_z(z)},\;\;\;
 A^2(z,t)=\frac{c(t)}{B_z(z)},&\nonumber\\ g(z,t)=GB^2_z(z)/2A^2(z,t),\;\;\; \tau(z,t)=\frac{1}{2}\frac{c_t(t)}{c(t)}.\nonumber \label{nonlinearity}
 \\
\end{eqnarray}
Here, $a_1$ signifies the centre-of-mass position of the condensate at the initial time, and $c(t)$ is an arbitrary positive definite function of time, which is taken as a constant, $c(t)=c$, in our illustration to account for zero loss/gain i.e., $\tau(z,t)=0$. The above equations have the direct dependence of phase and nonlinearity on the spatial derivative of the travelling coordinate $B_z(z)$ of the system, which will be decided by the external trap. We aim to find the exact form of the travelling coordinate $Z(z-vt)$ to find the explicit expression of all the related physical parameters by simplifying the above expressions along with Eq.~(\ref{consistency for potential}). We arrive at the expression for the spatial function of the travelling coordinate ($B(z)$), which helps us write the travelling coordinate as
\begin{equation}\label{zz}
    Z(z-vt)=\gamma e^{a_2}\int_{0}^{z}e^{\beta \cos(lz')}dz'-vt+a_1,
\end{equation}
where the first term is the function $B(z)$ and $a_1$, $a_2$ are constants of integration. The phase of the wavefunction is obtained by combining Eq.\ref{consistency for potential} and Eq.\ref{consistency for travelling coordinate} by using the form of $B(z)$ and $D(t)$:
\begin{eqnarray}\label{phase}
    \theta(z,t)=\frac{v}{\gamma}e^{-a_2}\int_{0}^{z}e^{-\beta \cos(lz')}dz'+\frac{\beta^2 l^2}{16}t+a_3.
\end{eqnarray}
Here, $a_3$ is an integration constant signifying an initial phase at the origin if there is any. One can also write the amplitude $A(z,t)$ and the nonlinearity $g(z,t)$ by substituting the evaluated expression of $B(z)$. The most crucial expression as far as the experimental utility is concerned is the physical parameters of the constituent lasers forming the BOL, and we obtain the potential parameters as
\begin{equation}
V_1=\frac{-\beta^2l^2}{16}, \;\;\; V_2=\frac{\beta l^2}{4}.
\end{equation}

These suggest that any arbitrary potential parameters may not support to excite solitary waves in BEC. Our analytical model contributes to that issue by identifying the parameter related to obtaining a solitary wave that can manifest Anderson localization. In fact, the ratio between the laser amplitudes, ${\frac{V_{1}}{V_{2}}}$, is obtained as ${\frac{\beta}{4}}$. Hence, ${\beta}$ controls the shape of the potential by relative changes in the laser amplitudes. This helps to manufacture an appropriate trap to investigate a variety of solitary excitation. We depict the potential profiles in Fig. \ref{fig:potential} to show the preparation of the trap for Anderson localization with higher $\beta$ for a given $l$. The geometrical lattice frustration appears and becomes dominant for larger $\beta$. The difference between lattice depth and frustrated depth is $\Delta d=\frac{\beta l^2}{2}$, which is responsible for the localization at higher ${\beta}$ values, as delineated by the condensate density in Fig. \ref{fig:potential}(b). This depth difference, responsible for localization, is similar to the depth difference in the tight-binding model, originally considered by Anderson for electronic systems \cite{anderson1958absence}. To analytically establish such localization, we evaluate all the functions required to express a time-dependent wavefunction of the condensate under BOL trap. It is well known that attractive inter-atomic interaction manifests bright solitary waves, whereas repulsive interaction supports dark solitary waves \cite{adhikari2009localization,atre2006class,burger1999dark,strecker2003bright}. It is already mentioned that a nonlinear Schr\"{o}dinger equation (NLSE) possesses solutions in the form of Jacobian elliptic functions; $cn[Z(z-vt),m]$ and $sn[Z(z-vt),m]$ for attractive and repulsive interactions, respectively. Accordingly, we can write the final form of the wavefunctions with appropriate elliptic functions:
\begin{widetext}
\begin{eqnarray}\label{wave}
    \psi(z,t)=\sqrt{\frac{c(t)}{\gamma e^{\beta \cos(lz)}}}cn\Bigg[(\gamma e^{a_2}\int_{0}^{z}e^{\beta \cos(lz')}dz'-vt+a_1), m\Bigg]e^{i\theta(z,t)},\\
    \psi(z,t)=\sqrt{\frac{c(t)}{\gamma e^{\beta \cos(lz)}}}sn\Bigg[(\gamma e^{a_2}\int_{0}^{z}e^{\beta \cos(lz')}dz'-vt+a_1), m\Bigg]e^{i\theta(z,t)},
\end{eqnarray}
\end{widetext}
where the expression of $\theta(z,t)$ is given in Eq. (\ref{phase}). The first equation is for an attractive regime and the latter represents repulsive interaction. In both cases, a family of solutions is found for various modulus parameters $m$. In the present work, we are interested on the localization of the condensate and consider the solution with $cn[Z(z-vt),m=1]=$sech $Z(z-vt)$ for further investigation. 

It is also well-known that bright excitations under attractive interaction are stable upto certain critical number of atoms ($N_{c}$) \cite{ruprecht1995time,roberts2001controlled,bao2003numerical,cornish2006formation,khaykovich2002formation,gammal2001critical,gammal2002critical}. This critical number of atoms is evaluated by comparing the nonlinearity obtained in Eq. (\ref{nonlinearity}) and the general relation of nonlinearity with space distributed scattering length \cite{kumar2020mass}. Some obtained critical numbers are tabulated  in Table \ref{table1}, for instance. Two different scattering lengths are considered with their maximum values to obtain the limit. The trap asymmetry $\Lambda$ and $K$-values are crucial for evaluating $N_{c}$. The obtained nonlinearity in our model is space distributed and also $ g(z)=\sqrt{2\pi\Lambda}N_{c}a_s(z)/ a_{\perp}$. The nonlinearity attains its maximum value at the vicinity of Anderson localization, which can be further reduced to increase the critical number of atoms. The experimental viability of our model will rely on these critical numbers.
\begin{table*}[htpb]
    \begin{tabular}{ | c | c | c | c |  }
    \hline
        \boldmath \hspace{.3cm}$\Lambda=\frac{\omega_{z}}{\omega_{\perp}}$ \hspace{.3cm} & \hspace{.3cm} \boldmath$K=\frac{N_{c}|a_{s,max}|}{a_{\perp}}$ \hspace{.3cm} & \hspace{.3cm} \boldmath $N_{c}\; for\;|a_{s,max}|=0.21nm$ \hspace{.3cm} & \hspace{.3cm} \boldmath$N_{c}\;for\; |a_{s,max}|=0.15nm$ \hspace{.3cm} \\  \hline
        0.1 & 0.460 & 3120 & 4370 \\  \hline
        0.5 & 0.560 & 3800 & 5320  \\ \hline
    \end{tabular}
    \caption{Critical number of atoms at two scattering lengths, for a set of aspect ratios of longitudinal and transverse trap frequencies and their corresponding $K$-values. The values of $\gamma$ and $c$ are taken as $0.1$ and $10^{4}$, respectively.}
    \label{table1}
\end{table*}

\begin{figure*}[htpb]
\centering
\includegraphics[width=15cm]{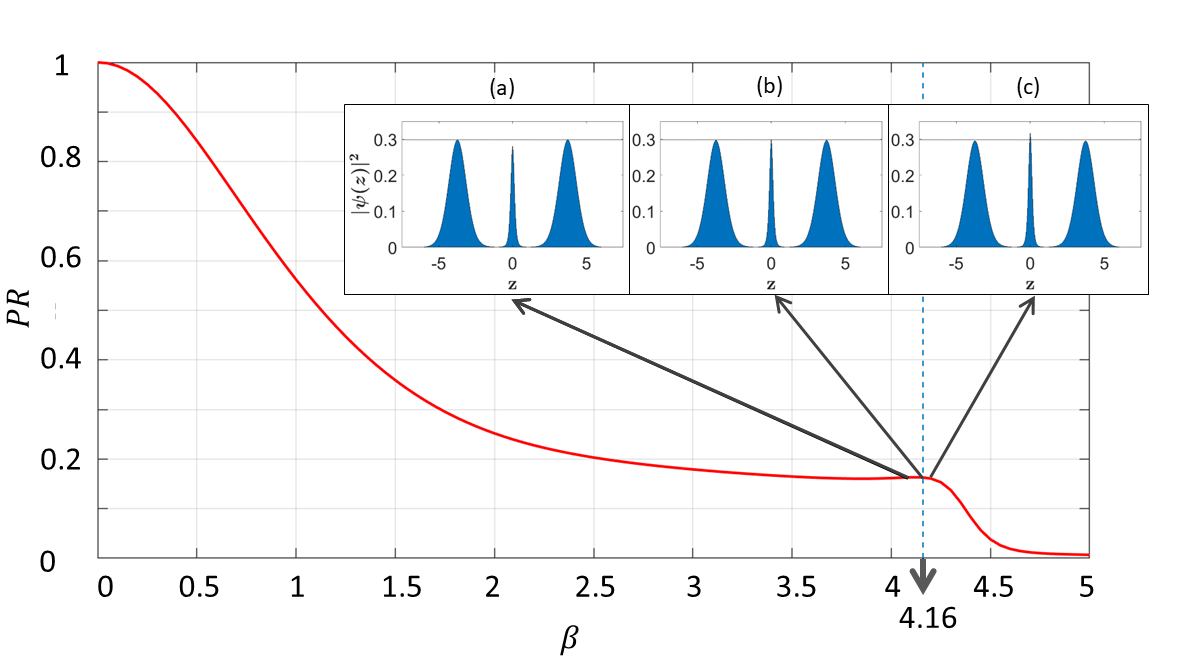}
\caption{Participation ratio for different ${\beta}$ values. The dashed line indicates $\beta_c=4.16$, RHS of which is the domain for Anderson localization. The inset shows the condensate densities for ${\beta}= a)\;4.15,\;b)\;4.16$ and $c)\;4.17$. Others parameters are given in dimensionless units: $G=-1$, $t=0$, $c=10^{4}$, $l=0.84$ and, $\gamma=0.1$.}
\label{fig:2}
\end{figure*}

\section{Identifying the Parameter Regime of Localization through Participation Ratio}
We are interested in the dynamics of Anderson localization, which should occur for attractive interactions. However, the solution covers various situations corresponding to the potential parameters, $\beta$ and, $l$, associated with the power and wavelength of the laser. For a given wavelength of the laser, we investigate the condensate cloud for various $\beta$ to identify the region of localization. The degree of localization is understood in terms of the participation ratio (PR) \cite{Lifshits1988IntroductionTT}, which  is defined in terms of the obtained wavefunction as \cite{mukherjee2015modulation},
\begin{equation}
    PR=\frac{1}{\int_{-\infty}^{\infty}|\psi(z,t)|^4 dz}.
\end{equation}
Physically, PR measures the number of lattice sites over which the wavefunction is extended. A PR value of $1$ denotes that the wavefunction extends over all the lattice sites, and a PR closer to zero denotes single-site participation. Figure \ref{fig:2} shows how PR changes with increasing the laser power. There is a smooth fall of PR for increasing ${\beta}$, and interestingly one could note a sudden descent towards zero after ${\beta}$ reaches $4.16$. This is an indication of the emergence of prominent localization from extended states. The value of ${\beta}$ at which this transition occurs is denoted by ${\beta_c}$. The inset in the Fig.\ref{fig:2} shows the condensate densities before, after, and at ${\beta_c}$. It is observed that at ${\beta}={\beta_c}$, the height of the density in the central site is equal to that in neighboring sites.
It is important to discuss the change in the behavior of the condensate for lower and higher ${\beta}$ \textit{w.r.t.} ${\beta_c}$. We have studied these two regimes by fitting them using suitable functions as shown in Eq.(\ref{PRfit}). And the parameters for the best fit are estimated as follows: $A_1=0.154, A_2=0.842, A_3=0.881, A_4=0.166$ for ${\beta<\beta_c}$ regime and $B_1=0.0086, B_2=0.152, B_3=17.83$ for ${\beta>\beta_c}$ regime:
\begin{eqnarray}
   PR_{(\beta<\beta_c)}=A_1+A_2e^{-A_3 \beta^2+A_4 \beta^3}\nonumber\\
   PR_{(\beta>\beta_c)}=B_1+B_2e^{-B_3(\beta-\beta_c)^2}.\label{PRfit}
\end{eqnarray}
\begin{figure}[!h]
    \centering
    \includegraphics[width=9cm]{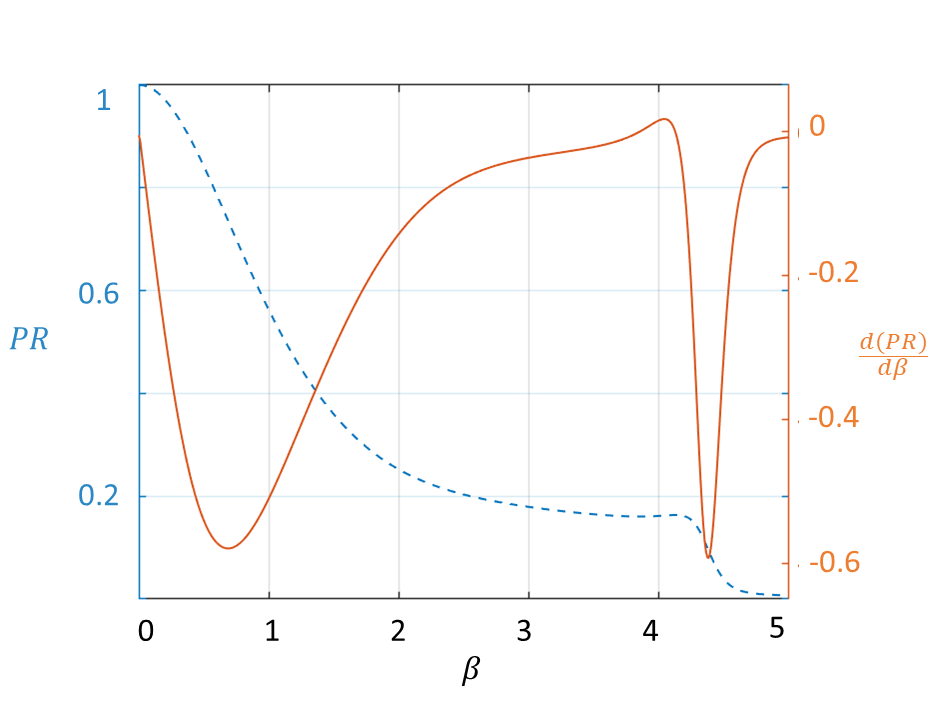}
    \caption{Participation ratio (dashed line) and its derivative {\it w.r.t.} ${\beta}$ (solid line) for various ${\beta}$ values. We have used the same parameters as in Fig.\ref{fig:2}.}
    \label{fig:3}
\end{figure}
It is worth observing that, for $\beta<\beta_c$, $A_3$ dominates over $A_4$. However, the widths of the Gaussian fit for both regions are quite dissimilar, such that $B_3 \gg A_3$. Hence, we infer that when the laser power is below the critical value ($\beta<\beta_c$), the rate at which the PR saturates is much slower than the rate at which it reaches saturation for $\beta>\beta_c$. The two regimes $\beta<\beta_c$ and $\beta>\beta_c$ are further understood by computing the slope of PR $\it{w.r.t}$ ${\beta}$. Figure \ref{fig:3} shows variation of $\frac{d (PR)}{d \beta}$ $\it{w.r.t.}$ $\beta$ along with the PR variation (dotted curve). The two minima indicate the constant variation of PR before it reaches saturation in the two regimes. The curves in the ${\beta>\beta_c}$ are much narrower than the first regime since the rate at which PR saturates is faster in the higher ${\beta}$ regime. The first regime implies the localization towards the first two neighboring lattice sites around the origin $z=0$. However, this is neither an exponential localization in the central frustrated site nor an Anderson localization. On the other hand, the second domain for ${\beta>\beta_c}$ manifests a distinct behavior as shown by the variations of PR and its slope. This emerges due to the localization in the central frustrated site when the frustrated lattice depth is significant compared to the main lattice depth. The transition point between these two domains is identified by the transition of the slope from positive to negative, which happens at $\beta=\beta_c=4.16$. The minimum of the curve in the second domain corresponds to $50\%$ occupancy of the central frustrated site around ${\beta}=4.48$. We will soon prove that the localization of the second domain is an exponential localization and is coined Anderson localization.
\begin{table*}[htpb]
    \begin{tabular}{ | c | c | c | c | c | }
    \hline
        \boldmath \hspace{.7cm}$\beta$ \hspace{.7cm} & \hspace{.7cm} \boldmath$a$ \hspace{.7cm} & \hspace{.7cm} \boldmath $b$ \hspace{.7cm} & \hspace{.7cm} \boldmath$\zeta$ \hspace{.7cm} & \hspace{.7cm}  \boldmath$R^2 value$ \hspace{.7cm} \\  \hline
        4.2 & -0.2021 & 9.4038 & 0.1274 & 0.9979 \\  \hline
        4.4 & -0.1171 & 12.4731 & 0.0938 & 0.9979 \\ \hline
        4.6 & -0.0687 & 16.7402 & 0.0707 & 0.9982 \\ \hline
        4.8 & -0.0401 & 22.6492 & 0.0541 & 0.9987 \\ \hline
        5 & 1 & 31.2252 & 0.0413 & 0.9991 \\ \hline
        5.2 & 1 & 42.7321 & 0.0321 & 0.9995 \\ \hline
        5.4 & 1 & 58.6585 & 0.0252 & 0.9997 \\ \hline
        5.6 & 1 & 80.1018 & 0.0198 & 0.9999 \\ \hline
        5.8 & 1 & 107.73 & 0.0158 & 0.9999 \\ \hline
        6 & 1 & 141.523 & 0.0127 & 0.9999 \\ \hline
    \end{tabular}
    \caption{The fitting parameters for the exponential localization: $a$, $b$ and, $\zeta$ of Eq.(\ref{expLoc}) with corresponding $R-squared$ values. The localization length is scaled by harmonic oscillator length $a_{\perp}=1.4245$ $\mu m$.}
    \label{table}
\end{table*}

\subsection{Characterising the Anderson Localization}
Before getting into the time-dependent dynamics, explicitly for the Anderson-localized cloud, we need to characterize it in terms of localization length to establish the exponential localization property. In disordered physical systems, it is observed that the localization falls exponentially in space as ${e^{\frac{-|z|}{\zeta}}}$, where ${\zeta}$ is called the Localization length \cite{ishii1973localization}. In our work, we consider the variation of the profile of the condensate cloud  \textit{w.r.t.} $z$ in the domain, demarcated by ${\beta>\beta_c}$. The variations are studied for many values of the potential parameter $\beta$. In each case, the curve is examined to fit the exponential nature:
\begin{equation}
    a+be^{-\frac{|z|}{\zeta}}. \label{expLoc}
\end{equation}

 \begin{figure}[!h]
    \includegraphics[width=9cm]{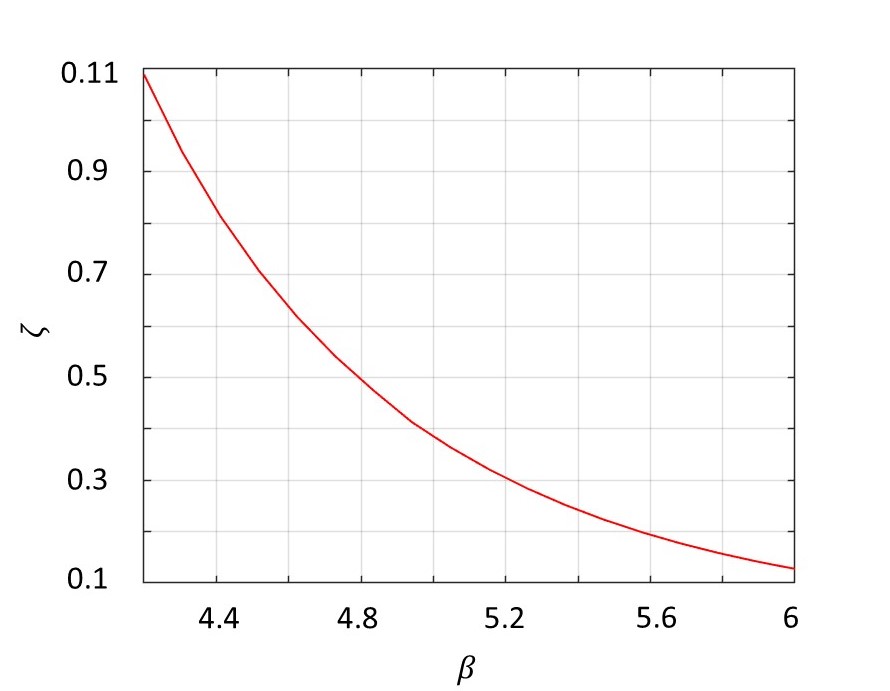}
    \caption{Variation of the localisation length $\zeta$ \textit{w.r.t.} $\beta$ in the localization domain, i.e., ${\beta>\beta_c}$. The other dimensionless parameters are $G=-1$, $t=0$, $c=10^{4}$, $l=0.84$ and $\gamma=0.1$. The localization length is scaled by harmonic oscillator length $a_{\perp}=1.4245$ $\mu m$.}
    \label{fig:4}
\end{figure}

We observe that the variations are indeed consistent with the exponential nature. The obtained localization length by curve fitting in $\zeta$ as per Eq.(\ref{expLoc}). The parameters for the best fit are estimated and are presented in Table \ref{table} with corresponding R-squared values. The R-squared values close to one indicate the goodness of our fit, confirming the existence of exponential localization or Anderson localization. On the other hand, the exponential function does not become a good fit for the condensate at ${\beta}<{\beta_c}$. The parameter $\zeta$ in the above fit gives us the localization length, which is obtained for laser powers ${\beta}>{\beta_c}$, and the results are shown in Fig. \ref{fig:4}. When one tries to control Anderson localization by altering the external trap, it becomes important to know the variation of the localization length with $\beta$. The length of localization falls steadily with increasing ${\beta}$, which also follows a regular pattern in the form of an exponential function:
\begin{equation}
    \zeta=c_1+c_2e^{-c_3 \beta}.\label{localization}
\end{equation}
The parameters $c_1$, $c_2$ and $c_3$ are evaluated as $0.0048$, $38.04$ and $1.406$, respectively. It is also observed that the localization length tends to saturate at $\zeta=0.0048$, implying a very sharp density spike for a very large value of $\beta$ without deviating from the exponential nature of the density profile.

\begin{figure}[!h]
    \centering
    \includegraphics[width=9cm]{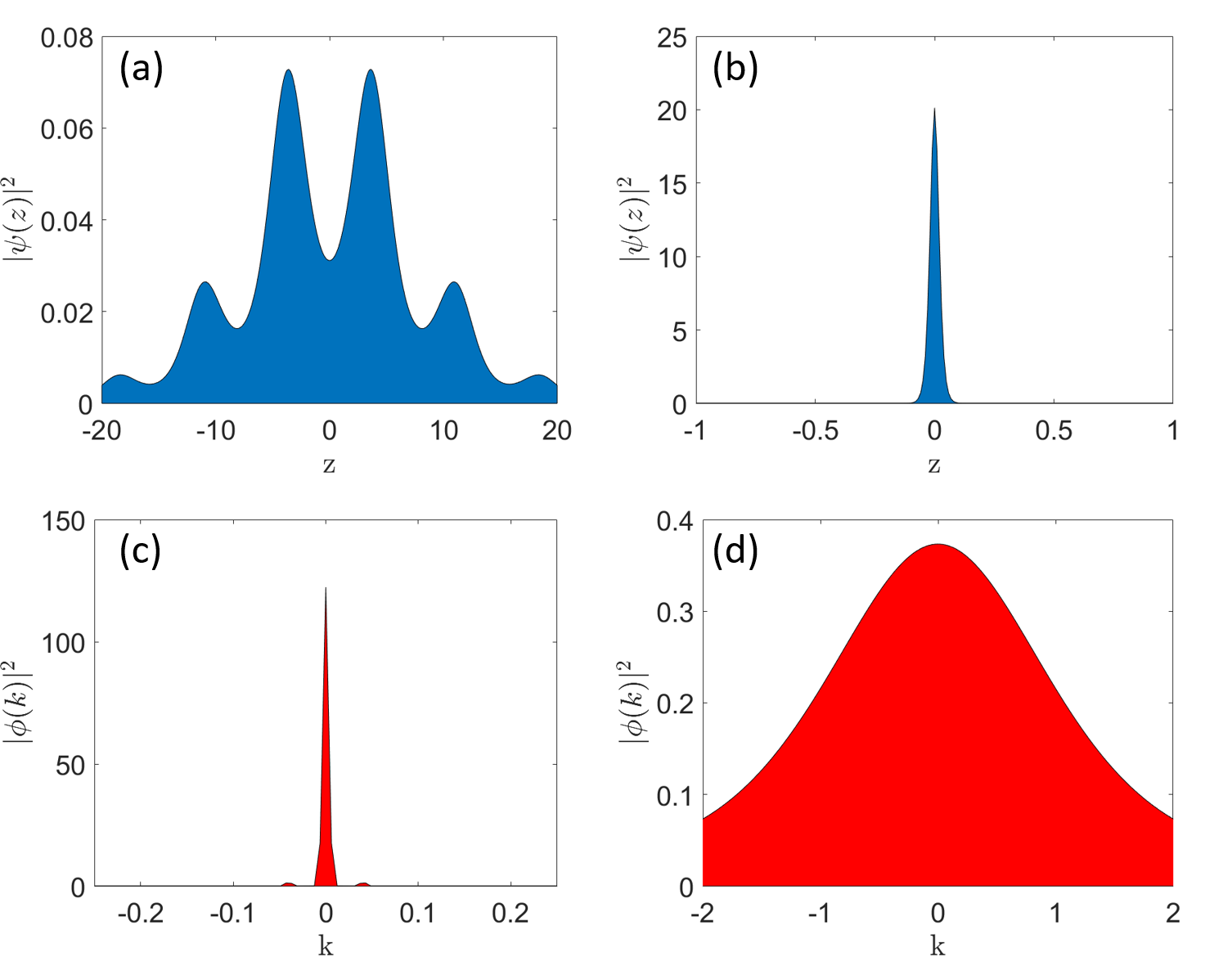}
    \caption{Condensate density in coordinate space for $\beta$ values a) $0.5$ and b) $6$. Condensate density in momentum space for $\beta$ values c) $0.5$ and d) $6$. Here, the dimensionless parameters include $G=-1$, $c=10^{4}$, $l=0.84$, $\gamma=0.1$ and $v=1$.  $k$ is scaled by the inverse of harmonic oscillator length $a^{-1}_{\perp}$.}
    \label{fig:3a}
\end{figure}

\begin{figure*}[htpb]
\centering
\includegraphics[width=15cm]{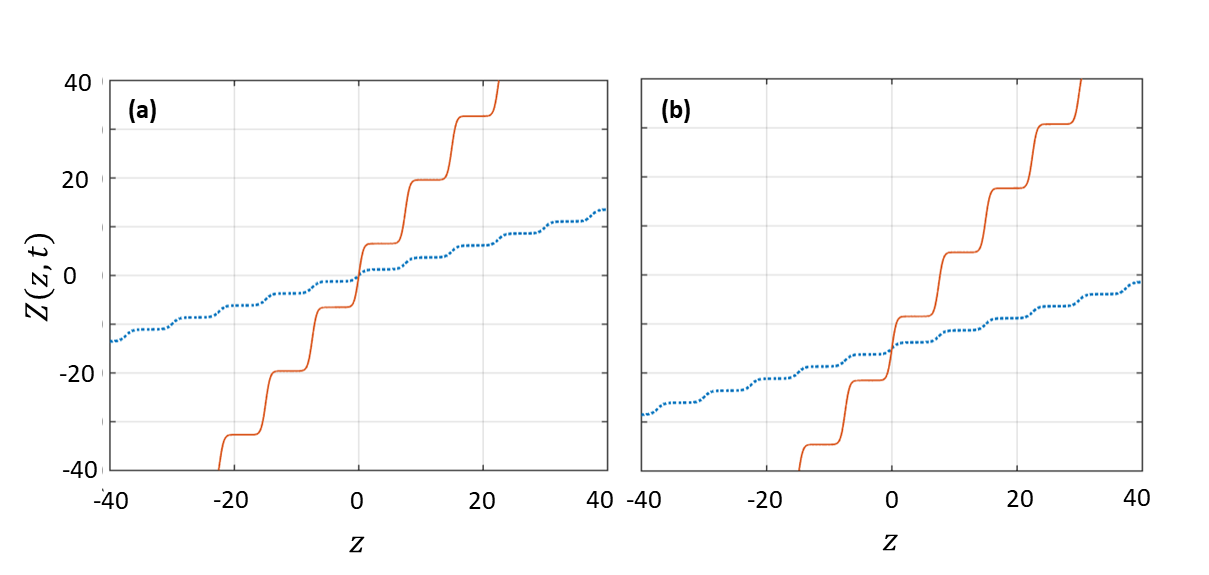}
\caption{Variation of travelling coordinate {\it w.r.t.} $z$ for times a) $t=0 ms$ and b) $t=13.5 ms$, and $\beta$ values $2.5$ (dotted line) and $4.5$ (solid line). Here, the dimensionless parameters include $G=-1$, $c=10^{4}$, $l=0.84$, $\gamma=0.1$ and $v=1$. The constants $a_1$ and $a_2$ are taken as zero for simplicity. $z$ is scaled by harmonic oscillator length $a_{\perp}=1.4245$ $\mu m$ and $t$ is scaled by the inverse of transverse frequency $\omega^{-1}_{\perp}=0.2242$ $ms$.}
\label{fig:1}
\end{figure*}

\subsection{Broadening of momentum distribution}
The comparison of condensate density in coordinate space with the momentum space gives us more insights into the physics of matter-wave localization. As it is expected from the uncertainty principle that the momentum distribution for the localized states must be broadened, indicating the range of momentum carried by the constituent atoms. Figure\ref{fig:3a}(a) and (b)  show the condensate density in coordinate space for $\beta=0.5$ and $6$ respectively, and Figure\ref{fig:3a}(c) and (d)  show the condensate density in momentum space for $\beta=0.5$ and $6$ respectively. For low disorders such as $\beta=0.5$, the BOL resembles an optical lattice with the BEC spread across various lattice sites corresponding to a single momentum peak in reciprocal space. This situation is reversed in higher disorders such as $\beta=6$, where the condensate is localized to a single site due to the scattering of component waves by frustrated lattice sites.

\section{Collapse and Revivals of Anderson Localization}
In the previous section, we have revealed the parameter domain for Anderson localization and how one can control it by tuning the external trap. Here, we will study the tunneling of the Anderson-localized cloud across the BOL sites.
The time-dependent density of the condensate is obtained from Eq.(\ref{wave}), which is a function of the travelling coordinate. Hence, knowing the nature of the travelling coordinate is important to analyze the tunneled cloud at real position $z$ and time $t$. Variations of $Z(z-vt)$ in Eq.(\ref{zz}) with real position coordinate at times, $t=0$ and $t=13.5$ ($\omega^{-1}_\perp$ units), are depicted in Fig. \ref{fig:1}(a) and (b), respectively. Each plot is shown for two $\beta$ values, one below $\beta_c$ ($2.5$) and another above $\beta_c$ ($4.5$). We observe that, for $\beta \rightarrow 0$, the travelling coordinate resembles a straight line of slope $\gamma$ for all times. However, higher $\beta$ values resemble a staircase-like travelling coordinate. The origin, $z=0$, does not depend on $\beta$ but creates an offset in the vertical axis due to the temporal drift of the condensate.
\begin{figure}[!ht]
\centering
    \includegraphics[width=8.5cm]{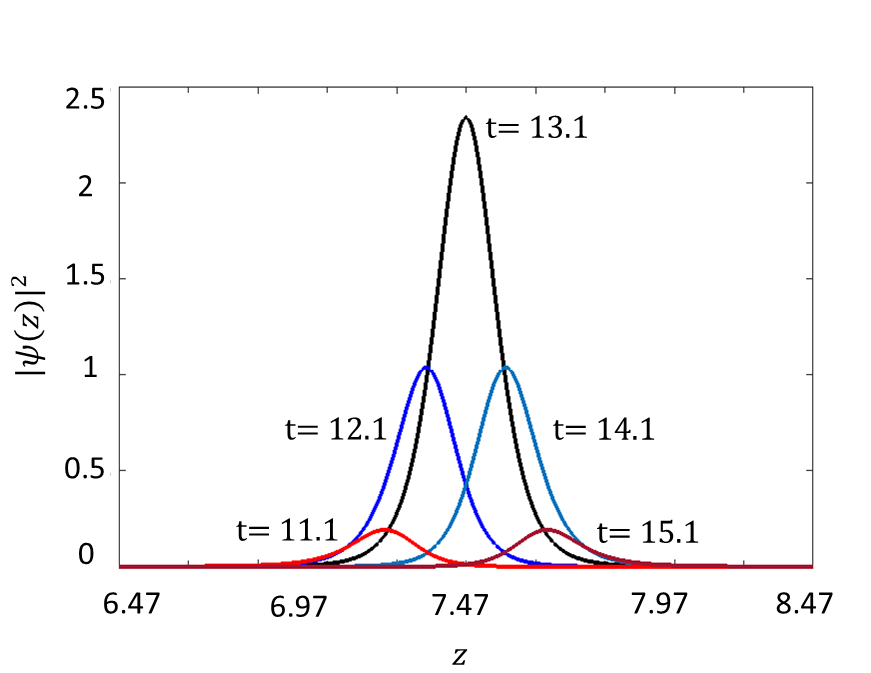}
    \caption{Time evolution of Anderson-localized condensate in the frustrated site for $\beta=4.5$. Here the dimensionless parameters includes $G=-1$, $c=10^{4}$, $l=0.84$, $v=1$ and $\gamma=0.1$. Here $z$ is scaled by harmonic oscillator length $a_{\perp}=1.4245$ $\mu m$ and $t$ is scaled by the inverse of transverse frequency $\omega^{-1}_{\perp}=0.2242$ $ms$.}
    \label{fig:6}
\end{figure}
In Fig.\ref{fig:6}, we closely observe the condensate density at the frustrated site ($z=7.47a_{\perp}$) in the domain of Anderson localization ($\beta=4.5$), we observe a rise and a decay of the localization with a small change in evolution time from $t=11.1$ to $t=15.1$ ($\omega^{-1}_\perp$ units). As we are dealing with a normalized density, the rise and decay of the localization signify that the cloud got tunneled in and tunneled out of the frustrated lattice site, respectively. Such tunneling of Anderson-localized cloud for a highly intense BOL trap (large $\beta$) seems interesting and encourages us to study the cloud post-decay with time.
 \begin{figure}[!h]
 \centering
  \includegraphics[width=8cm]{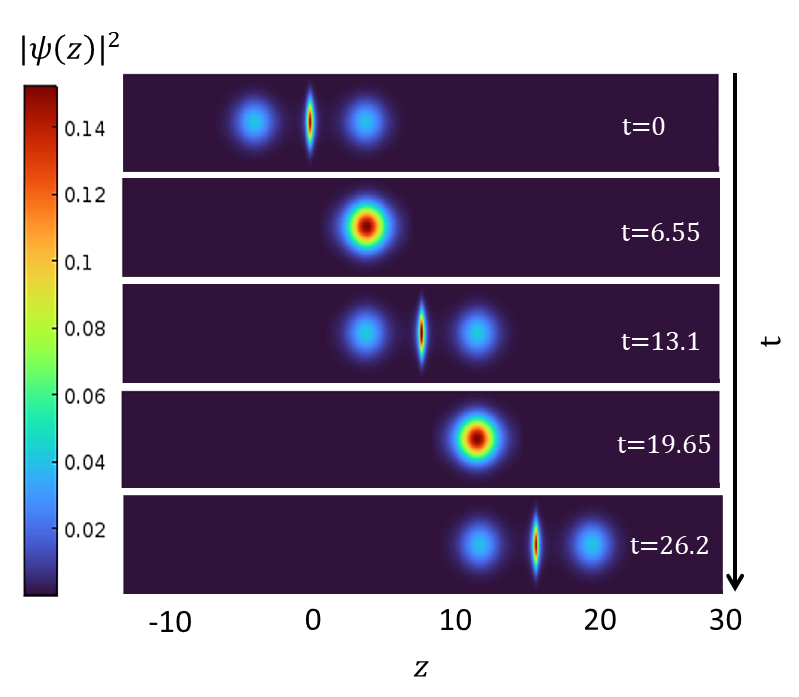}
\caption{Snapshots of the condensate density with $\beta=4.5$ at specific times; $0$, $t_R/2$, $t_R$, $3t_R/2$ and $2t_R$, where $t_R=13.1$. The dimensionless parameters being $G=-1$, $c=10^{4}$, $l=0.84$, $v=1$ and $\gamma=0.1$. Here $z$ and $t$ are scaled by harmonic oscillator length $a_{\perp}=1.4245$ $\mu m$ and inverse of transverse frequency $\omega^{-1}_{\perp}=0.2242$ $ms$ respectively, whereas $|\psi(z)|^2$ is in the unit of $a_{\perp}^{-1}$.}
\label{fig:5}
\end{figure}
\begin{figure*}[htpb]
\centering
\includegraphics[width=17.cm]{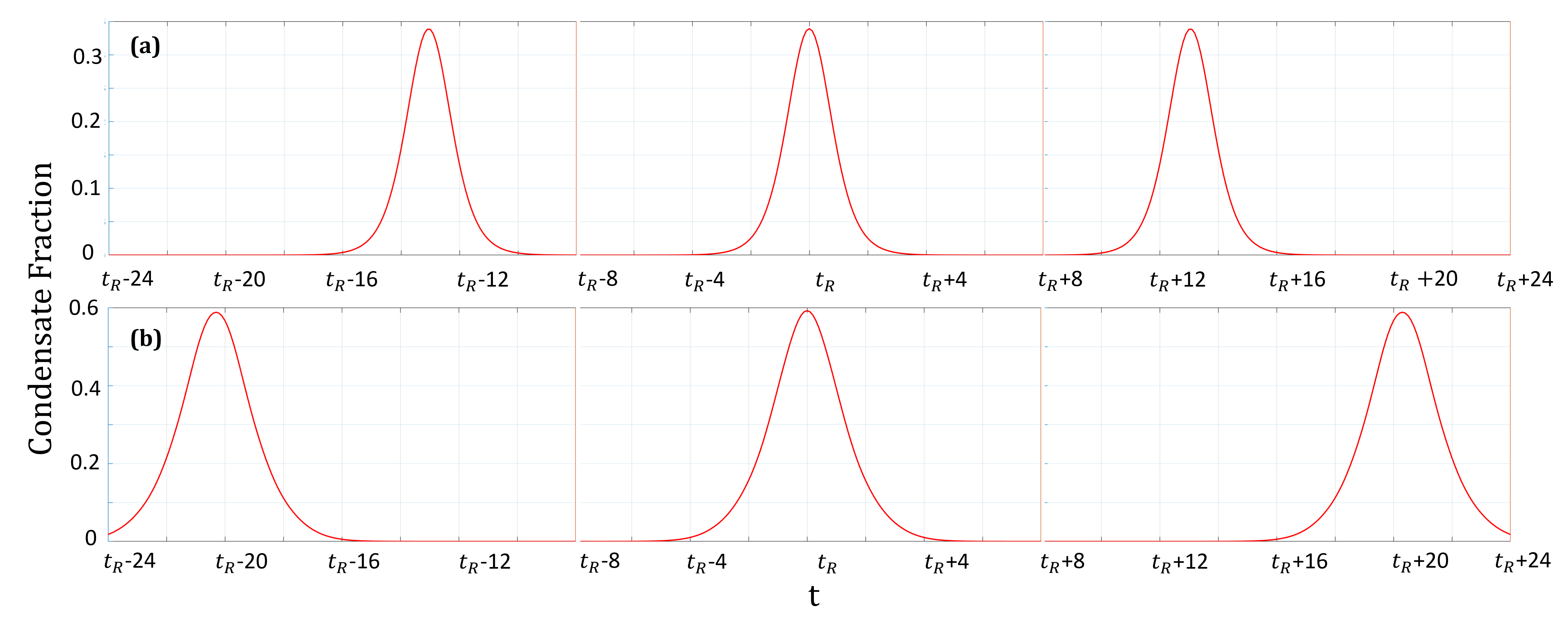}
\caption{Occurrence of Anderson localization with time for two ${\beta}$ values a) $4.5$, b) $5$. These are calculated by evaluating the condensate fraction at the frustrated site. The dimensionless parameters being $G=-1$, $c=10^{4}$, $l=0.84$, $v=1$ and $\gamma=0.1$. Here $t$ is scaled by the inverse of transverse frequency $\omega^{-1}_{\perp}=0.2242$ $ms$.}
\label{fig:7}
\end{figure*}

\subsection{The Revival Dynamics of Anderson Localization}

The rise and decay of the Anderson-localized condensate in the frustrated site is a clear signature of quantum mechanical tunneling across the lattice sites. However, Anderson localization happens at the frustrated lattice sites, which appear in spatially periodic intervals in a BOL trap. As each frustrated site is surrounded by two main lattice sites, the immediate tunneling of the density will happen to the neighboring lattice site instead of a frustrated one. In Fig. \ref{fig:5}, we have shown the time evolution of Anderson-like localization for $\beta=4.5$ and $l=0.84$. The nature of tunneling remains the same even for a very high value of $\beta$, where the condensate is just a sharp spike. Various density snapshots are shown at different times, $t=0,\;6.55, \;13.1, \; 19.65$ and $26.2$, in the unit of $\omega^{-1}_\perp$. One can observe that the Anderson localization in the frustrated lattice site revives after every $13.1 \omega^{-1}_\perp$, and the density becomes the replica of the initial Anderson-localized condensate. This time is designated as the revival time ($t_R$) of Anderson localization. During the half integrals of the revival time (i.e., $t=t_R/2,\;3t_R/2,...$), one could see an intermediate accumulation of the condensate density at the main lattice sites. Therefore, the Anderson-localized cloud not only tunnels across the lattice sites but also reappears at the next frustrated site after a time $t_R$. In general, this reoccurrence times are $t=nt_R$ for ($n=0,\;\pm1,\;\pm2,\;\pm3,..$), where $n=0$ corresponds to the central frustrated site.
\begin{figure}[!h]
\includegraphics[width=8.5cm]{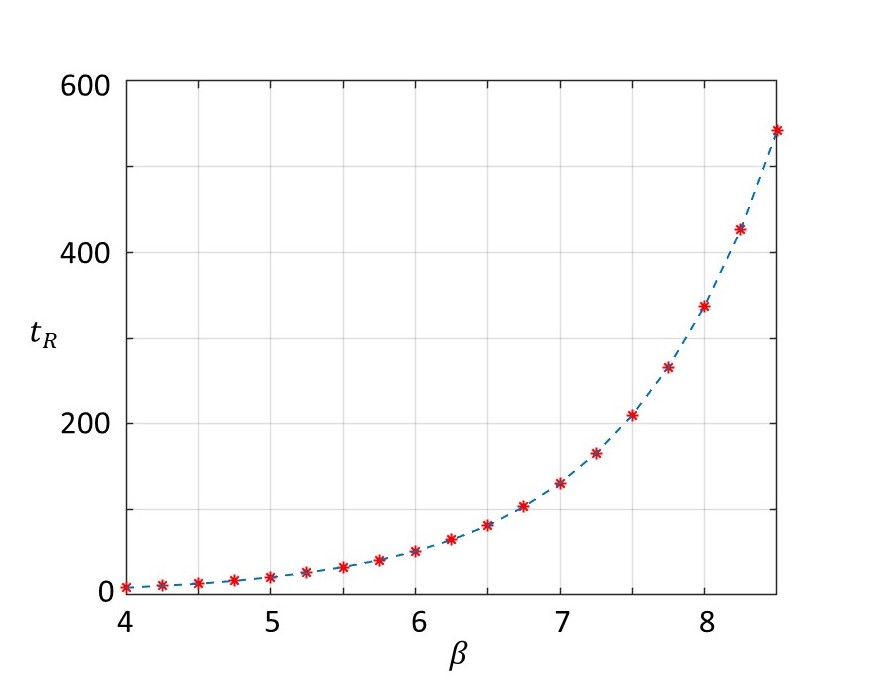}
    \caption{The variation of revival time ($t_R$) of Anderson localization with respect to ${\beta}$. $*$ symbols stand for evaluated points. The dimensionless parameters being $G=-1$, $c=10^{4}$, $l=0.84$, $v=1$ and $\gamma=0.1$. Here $t$ is scaled by the inverse of transverse frequency $\omega^{-1}_{\perp}=0.2242$ $ms$.}
    \label{fig:8}
\end{figure}
In Fig. \ref{fig:7}(a) and (b), we present the occurrences of Anderson-like localization \textit{w.r.t.} $t$ for $\beta=4.5$ and $\beta=5$, respectively. The occurrences are represented by the condensate fractions at the frustrated lattice sites with time relative to revival time, $t_{R}\pm t$. The maximum of each peak signifies a revival of Anderson localization, with the central peak corresponding to $t_{R}$. Two important aspects are worth noticing here: \textbf{i}) the widths of all the peaks are constant for a given $\beta$, and this constant width manifests the lifetime of the Anderson localization in any one of the frustrated lattice sites for an initially prepared BOL trap. Figure \ref{fig:7} suggests that the lifetime is greater for larger $\beta$ (Fig. \ref{fig:7}(b)), in spite of having a sharper Anderson localization. \textbf{ii}) the revival time (distance between two consecutive peak maxima) is different for different $\beta$. A larger $\beta$ or a sharper Anderson localization has a longer revival time. We also calculate the revival times for a tunable BOL (different $\beta$'s) and delineate the variation in Fig. \ref{fig:8} for the localization region ($\beta>\beta_c$). The revival time, $t_R$, increases exponentially with $\beta$ and takes a very high value for a very large $\beta$. This signifies a slowing down of tunneling with increasing $\beta$ and finally reaching a stationary Anderson localization of width $\zeta=c_1=0.0048$ (Eq.(\ref{localization})).

\begin{figure}[!h]
    \includegraphics[width=9cm]{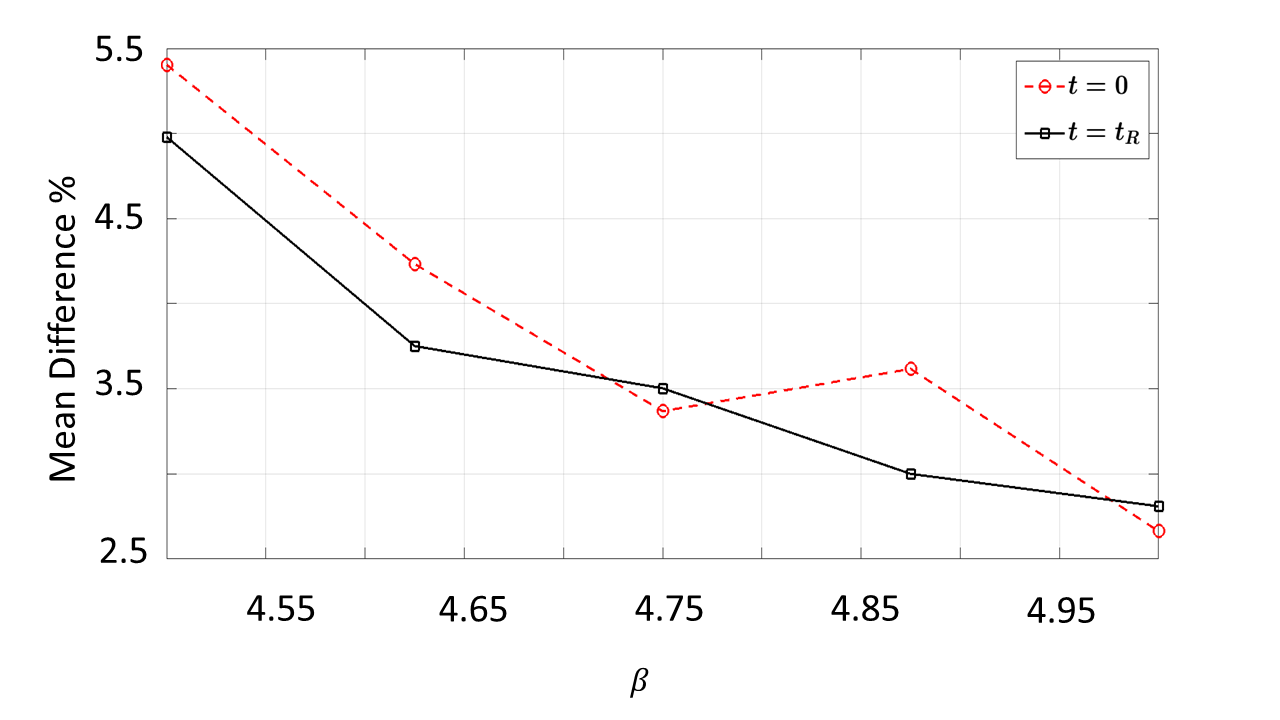}
    \caption{The variation of the mean difference of the condensate density with and without noise with respect to $\beta$. The black-line plot indicates the mean  difference at $t=t_R$, and the red-dashed plot indicates the mean difference at $t=0$.}
    \label{fig:9}
\end{figure}

\section{Dynamical and Structural Stability of the Anderson-localized Condensate}

\begin{figure}[!h]
    \includegraphics[width=9cm]{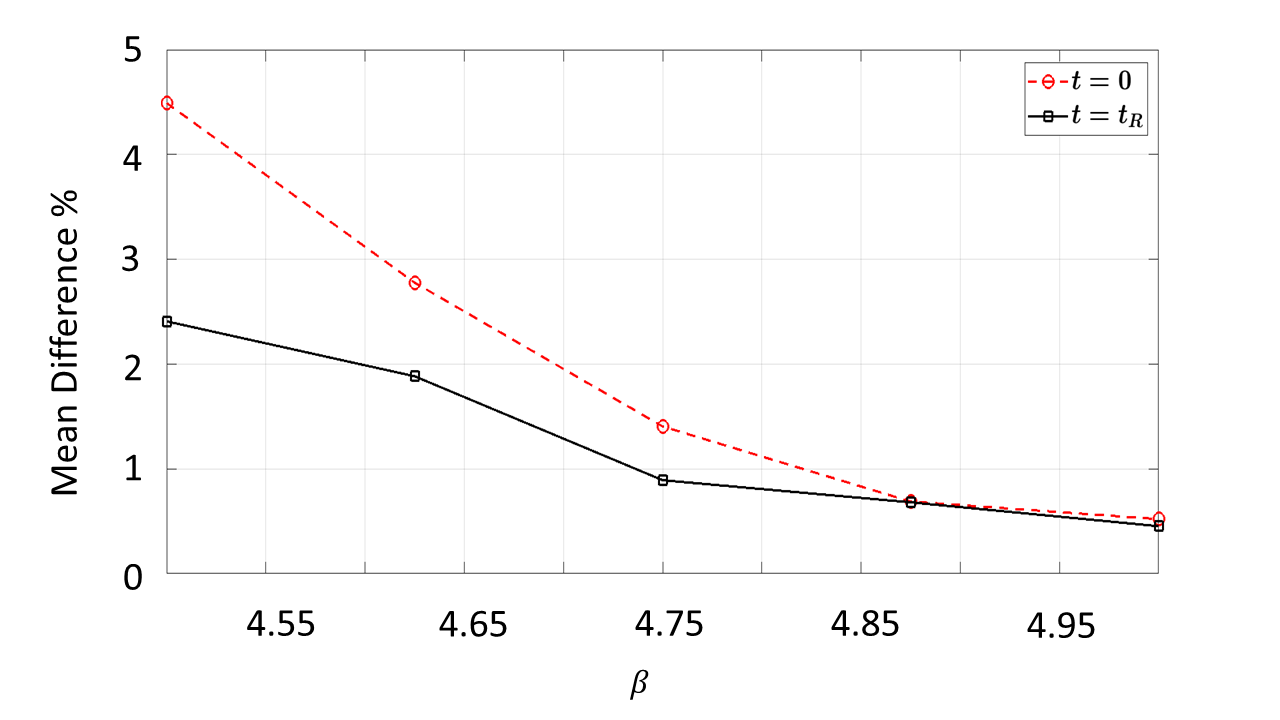}
    \caption{The variation of the mean difference of the condensate density with and without noisy potential with respect to $\beta$. The black-line plot indicates the maximum  difference at $t=t_R$, and the red-dashed plot indicates the maximum difference at $t=0$.}
    \label{fig:10}
\end{figure}

Anderson localization of BEC happens for a very strong BOL trap, which may hinder the condensate from being stable enough for experimental testing. Hence, we study the numerical stability of the Anderson-localized cloud for their occurrence at $t=0$ and $t=t_R$. These stability analyses are repeated for various BOL trap with different $\beta$ and are depicted in Fig. \ref{fig:9} and Fig. \ref{fig:10}. We perform a dynamical stability check, where the free evolution of the condensate with and without noise is compared at different times. We executed these analyses by numerically solving the GPE using the Split-Step Fourier method \cite{nath2020exact,bera2020matter,kundu2022synergy,halder2021exact}. The mean absolute difference of condensate density with noise from that without noise is estimated. We introduced a random noise $\Delta$, to our wavefunction $\psi(z,t)$ at times, $t=0$ and $t=t_R$:
\begin{eqnarray}
    \psi_{noise}(z,t=0)= \psi(z,t=0)+\Delta \nonumber\\
    \psi_{noise}(z,t=t_R)= \psi(z,t=t_R)+\Delta
\end{eqnarray}

In Fig.\ref{fig:9}, we show the mean absolute difference given by ${Mean[||\psi_{noise}(z,t)|^2-|\psi(z,t)|^2|]}$ with respect to $\beta$, which follows an decreasing trend. We have calculated the mean absolute difference for times $t=0$ and $t=t_R$, indicated by red-dashed and blue lines, respectively. It is interesting to find that at specific ${\beta}$ values, both the curves cross each other, thus making them equally stable. The most relevant point we observe is that the mean difference is always below $5\%$ for localization regime, even with a noise of $10\%$ added to the condensate.

In the case of structural stability analysis, the evolution of the condensate in the noise-added potential is compared with that in a noiseless potential. Figure \ref{fig:10} shows the mean absolute difference of condensate density in the presence and absence of random noise added to the potential for different $\beta$ values. The red-dashed and blue lines indicate the mean absolute difference for times $t=0$ and $t=t_R$. It is worth noting that the mean difference is below $5\%$ even for the deeper BOL trap ($\beta \geq 4.5$) with $10\%$ noise, considered in this work. Further, the cloud seems to be more robust to the fluctuations in potential compared to the fluctuations in the wavefunction. However, in both cases, the mean absolute difference decreases for higher $\beta$ values corresponding to the localization regime. Hence, the whole analysis presented in this work deals with a very stable Anderson-localized condensate under BOL.
\\
\section{Conclusion}

We have presented the time-varying solutions of 1D-GPE for stationary BOL with a special focus on the dynamics of localized solutions. We have discussed the nature of travelling coordinates and their role in the localization regime. The behavior of localization length and PR with respect to laser power indicates the emergence of strong localization. We prove that the localization that emerged inside the frustrated lattice site is exponentially varying in nature and is nothing but Anderson localization. The tunneling of Anderson localization from one site to another is thoroughly understood from the time snapshots of the condensate density. The dynamics of localization in the frustrated site are analyzed separately. The times of occurrence of localization are presented, and the revivals of the Anderson localized cloud are reported. The corresponding revival time is shown to increase with stronger BOL, implying a slowing down of the tunneling.  The dynamical stability for the initial time and first revival time are compared, and the values of ${\beta}$ for stable localization are reported. The work may be extended for other trap configurations and for a time-varying one.

\end{spacing}

\section{Acknowledgement}
PB acknowledges Prime Minister's Research Fellows (PMRF) Scheme for the financial support.

\bibliography{ref.bib}

\end{document}